\documentclass[aps,prapplied,twocolumn,showpacs,superscriptaddress,longbibliography]{revtex4-2}
\usepackage[utf8]{inputenc}
\usepackage[T1]{fontenc}
\usepackage{graphicx}
\usepackage{dcolumn}
\usepackage{amsmath}
\usepackage{amssymb}
\usepackage{textcomp}
\usepackage{verbatim}
\usepackage{float}
\usepackage{natbib}
\usepackage{hyperref}
\hypersetup{
    colorlinks=true,
    linkcolor=blue,
    filecolor=blue,
    urlcolor=blue,
    citecolor = blue,
}

\usepackage[symbol]{footmisc}
\usepackage{multirow}
\begin{document}
\title{Determination of carrier density and dynamics via magneto-electroluminescence spectroscopy in resonant tunneling diodes}
\author{E. R. Cardozo de Oliveira}
\affiliation{Departamento de Física, Universidade Federal de São Carlos, 13565-905, São Carlos, São Paulo, Brazil}
\author{A. Naranjo}
\affiliation{Departamento de Física, Universidade Federal de São Carlos, 13565-905, São Carlos, São Paulo, Brazil}
\author{A. Pfenning}
\email[]{ andreas.pfenning@physik.uni-wuerzburg.de}
\affiliation{Technische Physik, Physikalisches Institut and Röntgen Center for Complex Material Systems (RCCM), Universität Würzburg, Am Hubland, D-97074 Würzburg, Germany}
\author{V. Lopez-Richard}
\affiliation{Departamento de Física, Universidade Federal de São Carlos, 13565-905, São Carlos, São Paulo, Brazil}
\author{G. E. Marques}
\affiliation{Departamento de Física, Universidade Federal de São Carlos, 13565-905, São Carlos, São Paulo, Brazil}
\author{L. Worschech}
\affiliation{Technische Physik, Physikalisches Institut and Röntgen Center for Complex Material Systems (RCCM), Universität Würzburg, Am Hubland, D-97074 Würzburg, Germany}
\author{F. Hartmann}
\email[]{ fabian.hartmann@physik.uni-wuerzburg.de}
\affiliation{Technische Physik, Physikalisches Institut and Röntgen Center for Complex Material Systems (RCCM), Universität Würzburg, Am Hubland, D-97074 Würzburg, Germany}
\author{S. Höfling}
\affiliation{Technische Physik, Physikalisches Institut and Röntgen Center for Complex Material Systems (RCCM), Universität Würzburg, Am Hubland, D-97074 Würzburg, Germany}
\author{M. D. Teodoro}
\affiliation{Departamento de Física, Universidade Federal de São Carlos, 13565-905, São Carlos, São Paulo, Brazil}


\begin{abstract}
We study the magneto-transport and magneto-electroluminescence properties of purely n-doped GaAs/Al$_{0.6}$Ga$_{0.4}$As resonant tunneling diodes with an In$_{0.15}$Ga$_{0.85}$As quantum well and emitter prewell. Before the resonant current condition, magneto-transport measurements reveal charge carrier densities comparable for diodes with and without the emitter prewell. The Landau level splitting is observed in the electroluminescence emission from the emitter prewell, enabling the determination of the charge carrier build-up. Our findings show that magneto-electroluminescence spectroscopy techniques provide useful insights on the charge carrier dynamics in resonant tunneling diodes and is a versatile tool to complement magneto-transport techniques. This approach will drive the way for developing potentially more efficient opto-electronic resonant tunneling devices, by e.g., monitoring voltage dependent charge accumulation for improving built-in fields and hence to maximize photodetector efficiency and/or minimize optical losses.
\end{abstract}

\maketitle

\section{Introduction}

Resonant tunneling diodes (RTDs) with their current peak followed by a region of negative differential conductance (NDC)~\cite{Tsu,Tsu1,Sun98} are semiconductor devices with potential applications as terahertz oscillators~\cite{Feiginov,Feiginov2019}, fast switches~\cite{Growden2015}, and optically active elements in opto-electronic circuits~\cite{Siewe,Ironside2019} and photodetectors~\cite{Hartmann,Pfenning2016,Nie2019,Ahmadzadeh2020}. The charge carrier dynamics rules their functionalities and detection abilities. Given that charge excitation, accumulation, and transport control the device operational parameters and responsivity, it is important to study how all these three properties are combined. 

Charge carriers can be generated optically or injected by an applied voltage and the RTD response is affected by the way they are collected and stored. Magneto-transport measurements have been well-settled means of studying space charge build-up in RTDs as described in Refs. \onlinecite{Eaves1989,Fischer94,Leadbeater91,Eaves1988,Leadbeater1989}. However, as demonstrated in this work, blind spots can emerge where the magneto-transport tools are not able to provide a complete characterization of the charge accumulation scheme within the main operational range of the device. Under such circumstances it is necessary to appeal to a combination of both transport and optical observations as it is unveiled in the following discussion.

The insertion of an emitter prewell adjacent to the double barrier has proven to be a relevant design ingredient to study charge carrier accumulation.~\cite{Riechert1990,Choi92,Boykin99,Lewis2000,Pfenning2017,Pfenning2017_2} It improves the peak current density and the peak-to-valley current ratio (PVCR) at room temperature by suppressing states above the prewell, increasing the charge carrier density close to the emitter barrier and increasing the overlap of localized states in the prewell and quasi bound states in the double barrier quantum well (DBQW).~\cite{Riechert1990,Boykin99,Lewis2000} In the case presented in the current manuscript, the ability to clearly resolve the Landau quantization of the EL emission from states confined in the prewell, an effect unanticipated in previous studies, has been crucial. The resolution attained for this observation allows characterizing the filling of levels and subsequently, the charge accumulation. In order to do that, one must clearly ascribe first each emission line to the correct site at the RTD where the optical recombination occurs, as described in the text. In this sense, given the band structure tuning along the hetero-layered system, the spectroscopic tools are better suited for the charge build-up mapping along the growth direction. This is an advantage over just transport characterizations where the overlap of various simultaneous channels cannot be deconvoluted unlike for the clearly resolved emission lines of the optical spectrum.~\cite{Eaves1989,Fischer94,Yara2011} In unipolar doped RTDs, as it is the case under consideration here, EL appears by combining high electric fields at the lightly doped absorption region with relatively small gap that controls the impact ionization threshold~\cite{Cardozo2018,Pfenning1,White91,Hartmann2017} or potentially the Zener tunneling~\cite{Growden2018_2}. Thus, we can take advantage of the emitted light to investigate the charge carrier dynamics via a combination of both magneto-EL and magneto-transport. Although magneto-transport was extensively studied in the 1990s, and most works report the magneto-optic response through PL, the magneto-EL response was not previously explored in the way presented here. In this sense, we propose a complete investigation by complementing the transport properties and the intrinsic EL response along with the PL to describe the charge carrier dynamics in n-i-n GaAs/Al$_{0.6}$Ga$_{0.4}$As resonant tunneling diodes with In$_{0.15}$Ga$_{0.85}$As emitter prewells and compare with a conventional GaAs/AlGaAs RTD reference sample without prewell.

As the charge accumulation is the paramount working mechanism in RTD photodetectors, this approach for investigating these heterostructures will contribute on the development of potentially more efficient opto-electronic resonant tunneling devices, by e.g., monitoring voltage dependent charge accumulation for improving built-in fields and hence to maximize photodetector efficiency and/or minimize optical losses.

\section{Sample Design and Experimental Setup}

\subsection{Synthesis and structural characterization}

\begin{figure}[h!]
	\par
	\begin{center}
		\includegraphics[scale=0.4]{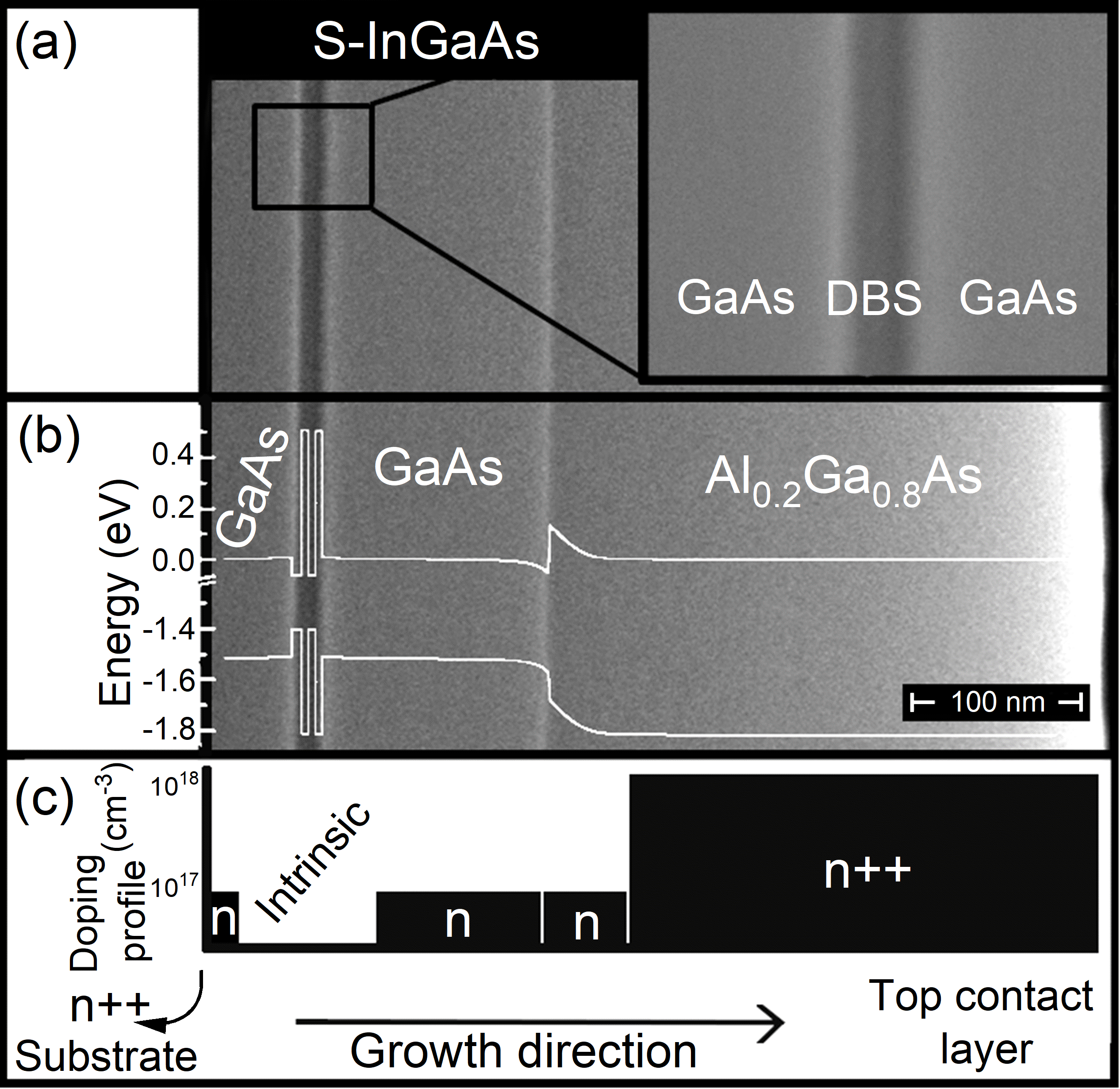}
	\end{center}
	\par
	\vspace{-0.5cm} \caption{(a) Cross-sectional scanning electron microscopy image of the S-InGaAs sample. A zoom on the double barrier structure is shown in the inset. (b) Bandstructure and (c) doping profile of the S-InGaAs RTD.}
	\label{SEM}
\end{figure}

Two samples were grown by molecular beam epitaxy. The first heterostructure consists of an intrinsic GaAs/Al$_{0.6}$Ga$_{0.4}$As double barrier structure (DBS), followed by a lightly doped ($n \approx 10^{17}$ $cm^{-3}$) 100 nm-thick GaAs drift region and a highly doped ($n \approx 10^{18}$ $cm^{-3}$) Al$_{0.2}$Ga$_{0.8}$As optical window. Growth details can be found in Ref. [\onlinecite{Cardozo2018}]. The InGaAs sample differs just in the intrinsic region where a 5 nm and 4 nm In$_{0.15}$Ga$_{0.85}$As pre- and quantum wells, respectively, were introduced. The sample with InGaAs pre- and quantum well is labeled as S-InGaAs, and the GaAs/AlGaAs RTD, used as reference, is labeled as Ref-GaAs. The  S-InGaAs RTD  layout and structural properties are shown in Fig.~\ref{SEM}. A cross-sectional scanning electron microscopy (SEM) image of the S-InGaAs RTD is displayed in Fig.~\ref{SEM}(a). Well defined interfaces can be observed. A zoom in the DBS (inset) exhibits the AlGaAs barriers (dark gray) and the InGaAs quantum well (light gray) between the barriers. The prewell, in the left side of the barriers, is barely seen due to the contrast between the layers. 

Figure~\ref{SEM}(b) presents the bandstructure, simulated at T = 4 K, according to the composition profile underlaid with the microscopy image. The matching between structural dimensions and the simulation indicates precisely controlled growth parameters. An accumulation region for electrons and a drift region for holes is present at the interface between GaAs and Al$_{0.2}$Ga$_{0.8}$As. The heterostructure doping profile is shown in Fig.~\ref{SEM}(c). Lightly and highly doped regions are indicated by the symbols n and n++, respectively, which corresponds to respective doping concentrations on the order of $10^{17}$ $cm^{-3}$ and $10^{18}$ $cm^{-3}$. The buffer layer, closer to the substrate, is not present in this image range, therefore, it is indicated by an arrow, with the corresponding donor concentration.

\subsection{Optical characterization}

\begin{figure}[h!]
	\par
	\begin{center}
		\includegraphics[scale=0.4]{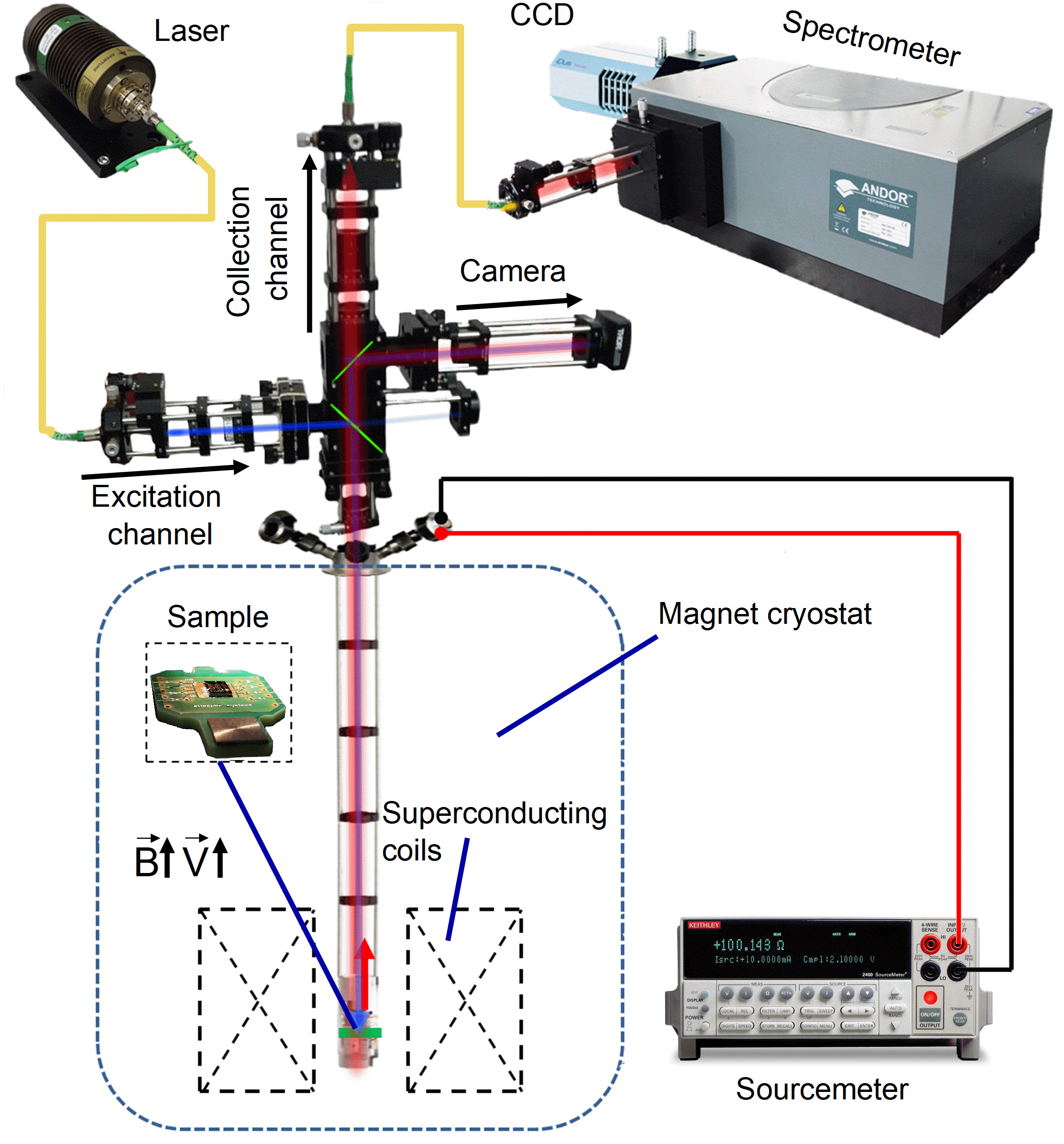}
	\end{center}
	\par
	\vspace{-0.5cm} \caption{Experimental setup for the electroluminescence and photoluminescence measurements (adapted from [\onlinecite{thesisCardozo}]).}
	\label{experimental setup}
\end{figure}

Electrical and optical measurements, with and without magnetic field, were performed with the sample placed inside a helium closed-cycle cryostat with superconducting magnet coils (Attocube - Attodry1000) and the magnetic field oriented parallel to the growth direction. For each magnetic field value, a voltage sweep was performed, and the EL signal and current were measured. The experimental setup is displayed in Fig.~\ref{experimental setup}. All measurements presented in this study were obtained at a nominal temperature of T = $4$ K. The optical signal is collimated by an aspheric lens (NA = 0.64) and transmitted along a 50 $\mu$m multimode optical fiber, being dispersed by a 75 cm spectrometer and detected by a silicon charge-coupled device detector (Andor - Shamrock/Idus).

\section{Results and Discussion}

\subsection{Optical and electrical properties at zero magnetic field}

\begin{figure}[h!]
	\par
	\begin{center}
		\includegraphics[scale=0.32]{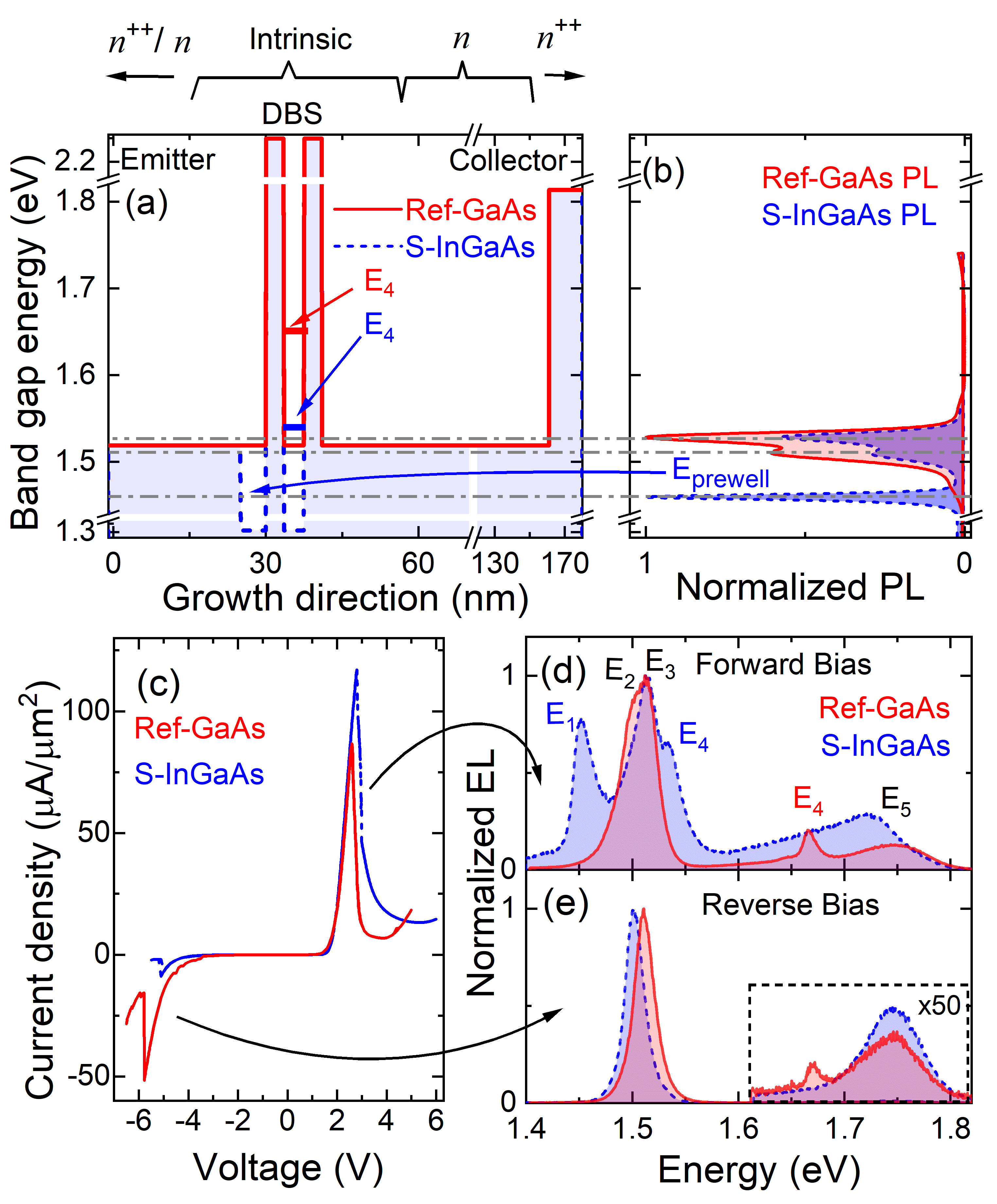}
	\end{center}
	\par
	\vspace{-0.5cm} \caption{(a) Bandgap energy profile of the S-InGaAs sample (blue-dashed line) and Ref-GaAs (red line). Bulk GaAs and donor levels are depicted as horizontal dashed lines. The DBQW confinement energies (E$_{4}$, see Table~\ref{table_peaks}) for Ref-GaAs and S-InGaAs are represented as red and blue lines, respectively. $n$ and $n^{++}$ indicate regions where the structure is lightly ($n\sim10^{17}$) and highly ($n\sim10^{18}$) doped, respectively. (b) PL spectra obtained for each sample in its respective color. Both samples present two emission lines related to donor and the bulk GaAs emissions. The S-InGaAs has an additional emission peak from the prewell. (c) Current-voltage characteristics for both samples. Arrows indicate the voltage biases where the EL signal was obtained at resonances. S-InGaAs (blue) and Ref-GaAs (red) EL spectra at (d) forward and (e) reverse bias voltage. Five emission lines are identified from $E_{1}$ to $E_{5}$ (see Table~\ref{table_peaks}).}
	\label{bandgap}
\end{figure}

The bandgap energy profiles at T = 4 K, of the two samples under study, are shown in Fig.~\ref{bandgap}(a). The normalized PL spectra obtained at 4 K and without applied voltage are presented in Fig.~\ref{bandgap}(b). For both samples, two emission lines are present, corresponding to the donor level~\cite{bogardus68,wagner87} and the bulk GaAs recombination, labeled as $E_{2}$ and $E_{3}$ on Table~\ref{table_peaks}, respectively. Both emission lines are almost identical, which indicates that the growth details (except of the prewell and QW) are nearly identical. For the S-InGaAs sample, the prewell emission ($E_{1}$ in Table~\ref{table_peaks}) can also be seen. Its emission line is more pronounced compared to the GaAs peak, with a peak height $\sim74\%$ higher than $E_{3}$.

The current density-voltage characteristics, $j(V)$, for Ref-GaAs and S-InGaAs are presented in Fig.~\ref{bandgap}(c). For forward bias voltage, the S-InGaAs (Ref-GaAs) peak current density is $j_{\mathrm{res}}= 117$ $\mu$A/$\mu$m$^{2}$ ($j_{\mathrm{res}}= 87$ $\mu$A/$\mu$m$^{2}$) at $V_{\mathrm{res}}= 2.8$ V ($V_{\mathrm{res}}= 2.6$ V), and the PVCR is 8.9 (12.8). As the heterostructure layouts (apart from the pre and quantum well) are identical, the resonant currents and voltages are comparable. After resonance voltage, the S-InGaAs valley current density comprises a longer voltage range with a minimum at $V = 5.3$ V, whereas the Ref-GaAs minimum valley current density is at $V = 3.8$ V. Moreover, the valley current density is higher for the S-InGaAs sample leading to a lower PVCR at 4 K. At reverse bias voltage, the S-InGaAs peak current density is $j_{\mathrm{res}}= -8.9$ $\mu$A/$\mu$m$^{2}$ at $V_{\mathrm{res}}= -5.10$ V, while the Ref-GaAs peak current density is $j_{\mathrm{res}}= -51.6$ $\mu$A/$\mu$m$^{2}$ at $V_{\mathrm{res}}= -5.78$ V. The heterostructure asymmetry leads to a reduction in the absolute value of the peak current density compared to forward bias, and this reduction is more pronounced for the prewell-containing heterostructure (see Ref. [\onlinecite{Lee2018}]).

\begin{table}[]
	\caption{Energy peak position of the emission lines for Ref-GaAs and S-InGaAs spectra measured via PL and EL (Figs.~\ref{bandgap}(b) and \ref{bandgap}(d)). The emissions are labeled as E$_{1}$ = InGaAs prewell, E$_{2}$ = donor GaAs, E$_{3}$ = bulk GaAs, E$_{4}$ = DBQW and E$_{5}$ = bulk AlGaAs. All units are in eV.}
	\label{table_peaks}
	\begin{tabular}{ccclclclclcl}
		\hline
		\multirow{2}{*}{Sample}   & \multicolumn{1}{l}{\multirow{2}{*}{Technique}} & \multicolumn{2}{c}{\multirow{2}{*}{E$_{1}$}}    & \multicolumn{2}{c}{\multirow{2}{*}{E$_{2}$}}    & \multicolumn{2}{c}{\multirow{2}{*}{E$_{3}$}}    & \multicolumn{2}{c}{\multirow{2}{*}{E$_{4}$}}    & \multicolumn{2}{c}{\multirow{2}{*}{E$_{5}$}}    \\
		& \multicolumn{1}{l}{}                           & \multicolumn{2}{c}{}                       & \multicolumn{2}{c}{}                       & \multicolumn{2}{c}{}                       & \multicolumn{2}{c}{}                       & \multicolumn{2}{c}{}                       \\ \hline
		\multirow{4}{*}{Ref-GaAs} & \multirow{2}{*}{PL}                            & \multicolumn{2}{c}{\multirow{2}{*}{--}}    & \multicolumn{2}{c}{\multirow{2}{*}{1.510}} & \multicolumn{2}{c}{\multirow{2}{*}{1.528}} & \multicolumn{2}{c}{\multirow{2}{*}{--}}    & \multicolumn{2}{c}{\multirow{2}{*}{--}}    \\
		&                                                & \multicolumn{2}{c}{}                       & \multicolumn{2}{c}{}                       & \multicolumn{2}{c}{}                       & \multicolumn{2}{c}{}                       & \multicolumn{2}{c}{}                       \\
		& \multirow{2}{*}{EL}                            & \multicolumn{2}{c}{\multirow{2}{*}{--}}    & \multicolumn{2}{c}{\multirow{2}{*}{1.497}} & \multicolumn{2}{c}{\multirow{2}{*}{1.514}} & \multicolumn{2}{c}{\multirow{2}{*}{1.660}} & \multicolumn{2}{c}{\multirow{2}{*}{1.760}} \\
		&                                                & \multicolumn{2}{c}{}                       & \multicolumn{2}{c}{}                       & \multicolumn{2}{c}{}                       & \multicolumn{2}{c}{}                       & \multicolumn{2}{c}{}                       \\
		\multirow{4}{*}{S-InGaAs} & \multirow{2}{*}{PL}                            & \multicolumn{2}{c}{\multirow{2}{*}{1.460}} & \multicolumn{2}{c}{\multirow{2}{*}{1.510}} & \multicolumn{2}{c}{\multirow{2}{*}{1.529}} & \multicolumn{2}{c}{\multirow{2}{*}{--}}    & \multicolumn{2}{c}{\multirow{2}{*}{--}}    \\
		&                                                & \multicolumn{2}{c}{}                       & \multicolumn{2}{c}{}                       & \multicolumn{2}{c}{}                       & \multicolumn{2}{c}{}                       & \multicolumn{2}{c}{}                       \\
		& \multirow{2}{*}{EL}                            & \multicolumn{2}{c}{\multirow{2}{*}{     1.452     }} & \multicolumn{2}{c}{\multirow{2}{*}{     1.496     }} & \multicolumn{2}{c}{\multirow{2}{*}{     1.514     }} & \multicolumn{2}{c}{\multirow{2}{*}{     1.534     }} & \multicolumn{2}{c}{\multirow{2}{*}{     1.740     }} \\
		&                                                & \multicolumn{2}{c}{}                       & \multicolumn{2}{c}{}                       & \multicolumn{2}{c}{}                       & \multicolumn{2}{c}{}                       & \multicolumn{2}{c}{}                       \\ \hline
	\end{tabular}
\end{table}

After surpassing a critical voltage of $V\geq1.8$ V, the EL emission is observed.~\cite{Cardozo2018} To gain an accurate understanding about the EL origin, normalized EL spectra for both S-InGaAs and Ref-GaAs samples were obtained at resonant current conditions (EL at V$=$V$_{\mathrm{res}}$) and are shown in Figs.~\ref{bandgap}(d) and (e), for forward and reverse bias voltages, respectively. At forward bias, the S-InGaAs EL spectrum shows five main emission lines, according to the Table~\ref{table_peaks}. In turn, Ref-GaAs spectrum consists of four emission lines, without the lower energy peak ascribed to the prewell, as observed for the S-InGaAs sample. The emission lines E$_{1}$ for S-InGaAs, and E$_{2}$ and E$_{3}$ for both heterostructures are lower in energy compared to the PL, probably due to the Joule heating. The DBQW emission energy $E_{4}$ differs between the two samples. With EL measurements we have access to the quantum well states during the resonant tunneling while measuring the I-V. It is not possible to observe DBQW emission line through PL at V $=$ 0 V, whereas measuring PL with V $\neq$ 0 V, the incident light can disturb the system and change the intrinsic charge carrier dynamics and DBQW quantization due to the large photogeneration of electron-hole pairs.~\cite{Carvalho2006,LdosSantos2008,LdosSantos2011}  For the Ref-GaAs we can extract the sum of electron and hole quantization energies (E$_{\mathrm{e^{-}}}$ $+$ E$_{\mathrm{h}}$ $=146$ meV) by subtracting E$_{4}$ from E$_{3}$, whereas the same is not possible for S-InGaAs as we do not have optical information from the bottom of InGaAs QW.

At reverse bias, both S-InGaAs and Ref-GaAs EL spectra present the GaAs and AlGaAs emission lines, and the DBQW peak is visible only for the reference sample. Under reverse bias, electrons undergo impact ionization mostly in the highly doped GaAs layer at the substrate side. Generated holes drift through the DBS towards the top contact side and accumulate at the interface between the GaAs and AlGaAs valence band barrier. The presence of $E_{3}$ and the missing of donor emission, $E_{2}$, suggest electron-hole recombinations occurring along the lightly doped GaAs layer. A portion of holes overcome the barrier and reach the optical window. The absence of InGaAs prewell emission at reverse bias indicates that the prewell is completely depleted of electrons.

\begin{figure}[h!]
	\par
	\begin{center}
		\includegraphics[scale=0.31]{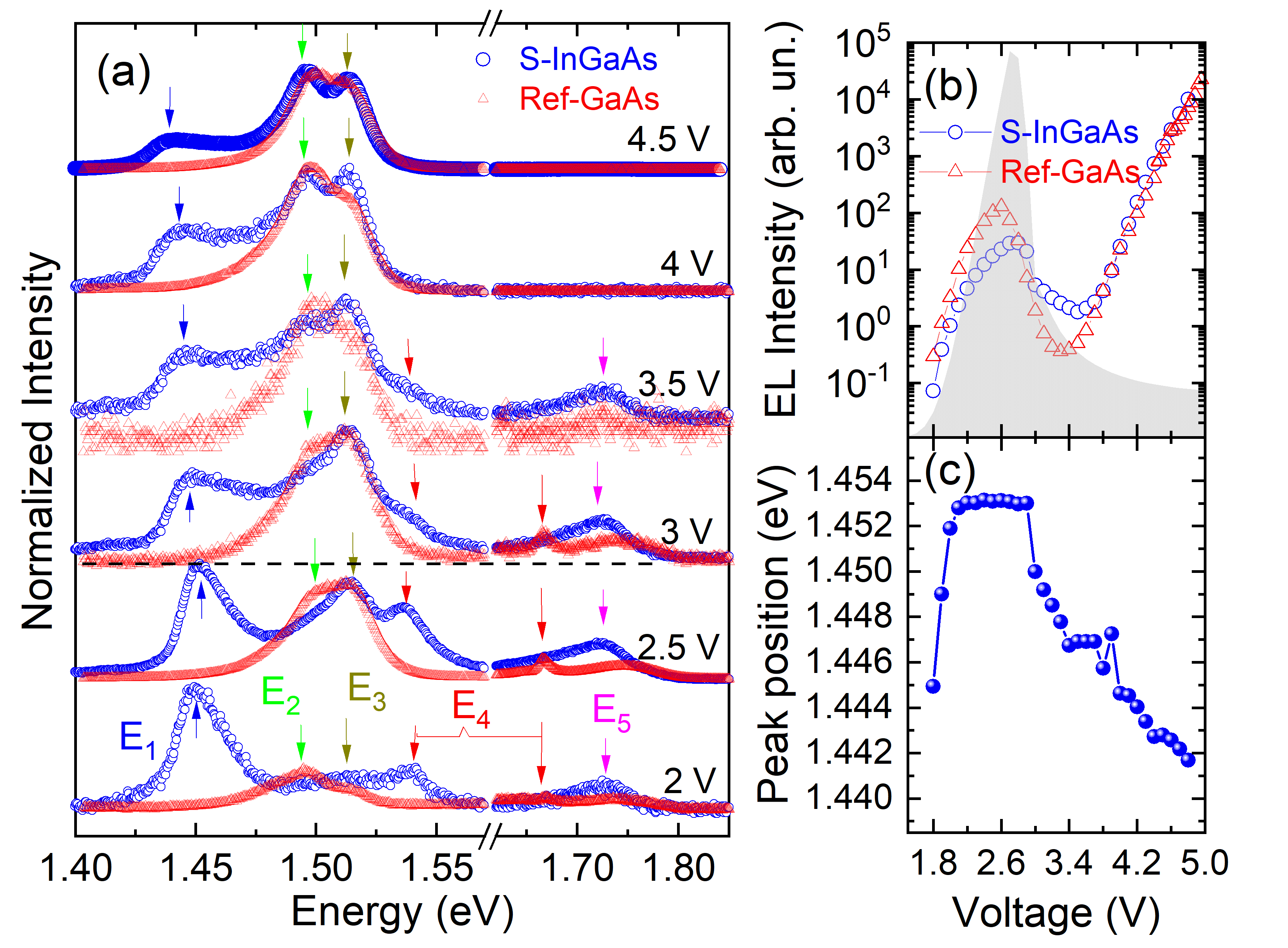}
	\end{center}
	\par
	\vspace{-0.5cm} \caption{(a) Normalized EL spectra for S-InGaAs (blue circles) and Ref-GaAs (red circles) for different bias voltages. Ref-GaAs emissions are normalized according to the maximum of S-InGaAs, disregarding the prewell emission. (b) S-InGaAs (blue opened circles) and Ref-GaAs (red opened triangles) integrated intensity vs. bias voltage. A gray shadow of the S-InGaAs I-V characteristics is also plotted. (c) Prewell peak position as function of voltage (blue dots).}
	\label{el_iv_xc_ratio}
\end{figure}

The EL emission across the whole voltage range provides insight into the internal charge carrier transport processes as well as the evolution of the band profiles. S-InGaAs and Ref-GaAs normalized EL spectra for different voltage values are presented in Fig.~\ref{el_iv_xc_ratio}(a).

The calculated impact ionization threshold is $E_{th}^{GaAs}= 1.80$~eV for the GaAs, and $E_{th}^{AlGaAs}= 2.08$~eV for the AlGaAs optical window, obtained by considering energy and momentum conservation in the transitions during ionization processes.~\cite{Hartmann2017,Redmer} For both heterostructures the onset for EL starts at V = 1.8 V due to $E_{th}^{GaAs}$. The AlGaAs peak, $E_{5}$, is present at voltages V $>$ 2.0 V, which means that a fraction of electrons travel ballistically through the drift region without impact ionization or scattering events before reaching the AlGaAs optical window, and, as this region is not well defined, EL onset variations may occur. As the applied electric field increases, more holes are created at the optical window, increasing the intensity of the emission $E_{5}$. At higher bias voltages (V $>$ 3.5 V), $E_{5}$ starts to vanish because generated holes at the AlGaAs region are swiped out towards the DBQW before they are able to recombine with electrons. The light emission from the DBQW, $E_{4}$, is also present at low bias voltages and increases up to the resonance condition as the quantum well gets populated. After the resonance voltage, $E_{4}$ emission abruptly drops because, in the off-resonance condition, the electron density inside the quantum well is significantly reduced.~\cite{VanHoof92}

At low bias voltages (2.0 V) the prewell emission, $E_{1}$, dominates. Holes created at the top-contact side drift towards the DBS and either recombine with electrons in the lightly doped (see Figs.~\ref{SEM}(b) and (c)) and/or intrinsic GaAs region (E$_{3}$), DBQW (E$_{4}$) or tunnel through the DBS. Here, they recombine with electrons in the prewell ($E_{1}$) and with electrons in the highly doped GaAs region of the emitter side ($E_{2}$). On the other hand, as the Ref-GaAs does not have a prewell, the holes recombine mostly with electrons from GaAs layers. When the bias voltage is above the resonance (V $\geq$ 3.0 V), electron charge build-up in the prewell is supported, which results in an increasing asymmetry of its EL emission ($E_{1}$). Furthermore, holes are more likely to be swiped into the highly doped region at the substrate side and recombine with electrons. Thus, the emission E$_{2}$ becomes more predominant compared to E$_{3}$.

Ref-GaAs and S-InGaAs EL integrated intensity vs. voltage are shown in Fig.~\ref{el_iv_xc_ratio}(b). A gray shadow of the S-InGaAs I-V characteristics is also plotted. Both intensity curves are comparable, with a peak at resonance, followed by an intensity drop in the valley region. At high voltages, the intensity increases again. Within the NDC region, the RTD without prewell presents an EL intensity decrease of almost three orders of magnitude. One order of magnitude reduction is observed for S-InGaAs. We have demonstrated, in a recent work, that higher EL peak-to-valley ratio (PVR) for Ref-GaAs is due to a competition between coherent and sequential tunneling channels.~\cite{Cardozo2018} The smaller optical PVR for the S-InGaAs heterostructure at cryogenic temperatures is probably caused by the prewell charge build-up. By analyzing the prewell peak position in Fig.~\ref{el_iv_xc_ratio}(c), obtained at the peak maximum intensity, we observe an increase from 1.445 eV to 1.453 eV before 2.0 V and then it becomes nearly constant up to the resonant voltage due to an electrostatic feedback, screening the prewell (see Refs. [\onlinecite{Fischer94}] and [\onlinecite{Leadbeater91}]). After the resonance, we can observe a constant redshift of the prewell energy of about $-4.40 \pm 0.24$ meV/V.

In this section we investigated the charge carrier dynamics as function of the applied voltage, at $B=0$~T, by analyzing the electroluminescence variations along the voltage range; and addressed the differences between the structures with and without the emitter prewell. The next section will be focused on the electrical and optical properties dependence with magnetic field and the determination of the charge carrier density by two different means.

\subsection{Magnetic field dependence of EL and electric transport}

\begin{figure*}[t!]
	\par
	\centering
	\begin{center}
		\includegraphics[scale=0.51]{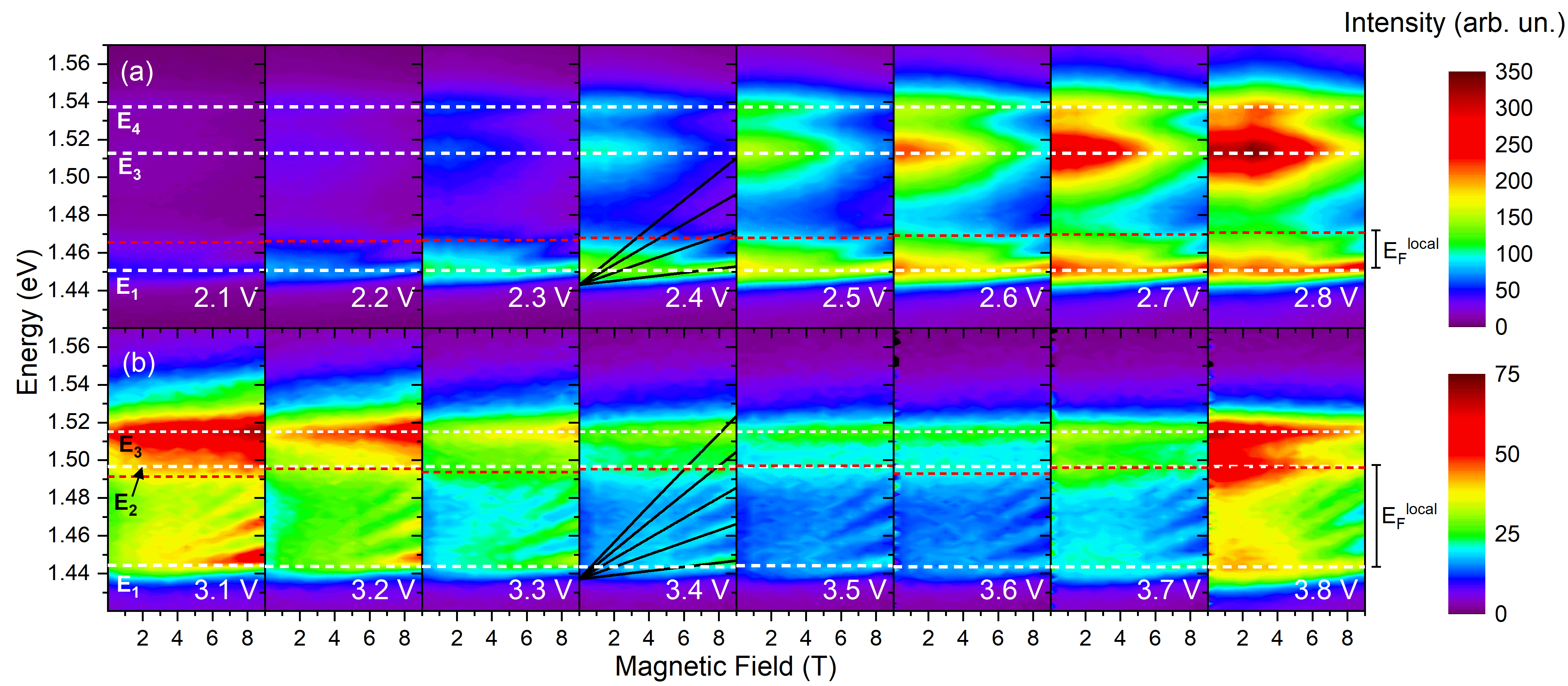}
	\end{center}
	\par
	\vspace{-0.5cm} \caption{Colormap of S-InGaAs EL signal vs. magnetic field for voltages (a) before (upper panels) and (b) after resonance (lower panels). Black lines in the middle graphs indicate the Landau levels splitting calculated with Eq.~\ref{LL}.}
	\label{landau}
\end{figure*}

Figures~\ref{landau}(a) and (b) present the two-dimensional intensity maps of S-InGaAs EL spectra as function of applied magnetic field and voltage. The top and bottom panels correspond to the measurements performed before and after the resonance current peak, respectively.

We can resolve a fan-like pattern of Landau levels (LLs) in the EL spreading out from the emitter prewell emission line as the magnetic field increases within a wide voltage range. On the other hand, no LL splitting is observed for the DBQW state. LLs are also absent from Ref-GaAs heterostructure (not shown). The differences of LLs splitting before and after the resonant condition are worth noting. At voltages up to the resonance, we observe a picture in which the levels are not well resolved. This indicates that the quasi-Fermi energy at the prewell, $E_{F}^{prewell}$, is close to the prewell ground state. In this case the DBQW is supporting a strong charge build-up due to resonant conditions and the electrons tunnel through the double barrier rather than accumulate in the prewell. On the off-resonance case, the energy levels misalignment quenches the resonant tunneling rate and electrons accumulate in the prewell, raising $E_{F}^{prewell}$ at higher energies.~\cite{Goodings94}

A peculiar feature is also observed at the voltages from 3.1 V to 3.5 V (Fig.~\ref{landau}(b)), where the EL intensity tends to increase with magnetic field, whereas at voltages below and above this voltage range the opposite occurs. The Ref-GaAs structure EL intensity at this voltage (not shown) increases with magnetic field as well.

From the experimental dependencies of the transition energies on the magnetic field, we calculate the energy separation of the LLs using a two-band model.~\cite{JMaan81,MPotemski90} In this model the energies of the Landau transitions between a parabolic band for the holes and the non-parabolic band for the electrons are given by the equation
\begin{multline}
E_{N}=\frac{E_{g}}{2}+\sqrt{\left(\frac{E_{g}}{2}+E_{0}\right)^{2}+E_{g}\left(N+\frac{1}{2}\right)\frac{\hbar eB}{m_{0,e}}} \\ +\left(N+\frac{1}{2}\right)\frac{\hbar eB}{m_{0,h}}+H_{0}
\label{LL}
\end{multline}
where $m_{0,e} = 0.071m_{0}$~\cite{EJones89} and $m_{0,h}=0.15m_{0}$~\cite{KLee2003} are the effective masses of the In$_{0.15}$Ga$_{0.85}$As conduction and valence bands, respectively; E$_{g}$ is the band gap energy of the In$_{0.15}$Ga$_{0.85}$As prewell; $N$ is the Landau level quantum number; $E_{0}$ and $H_{0}$ are the electron and hole subband energies. The calculation results are shown as solid lines in the colorplots of Fig.~\ref{landau} at 2.4 V and 3.4 V, where it can be seen that all Landau energies fit well with the experiment.

\begin{figure}[t!]
	\par
	\centering
	\begin{center}
		\includegraphics[scale=0.5]{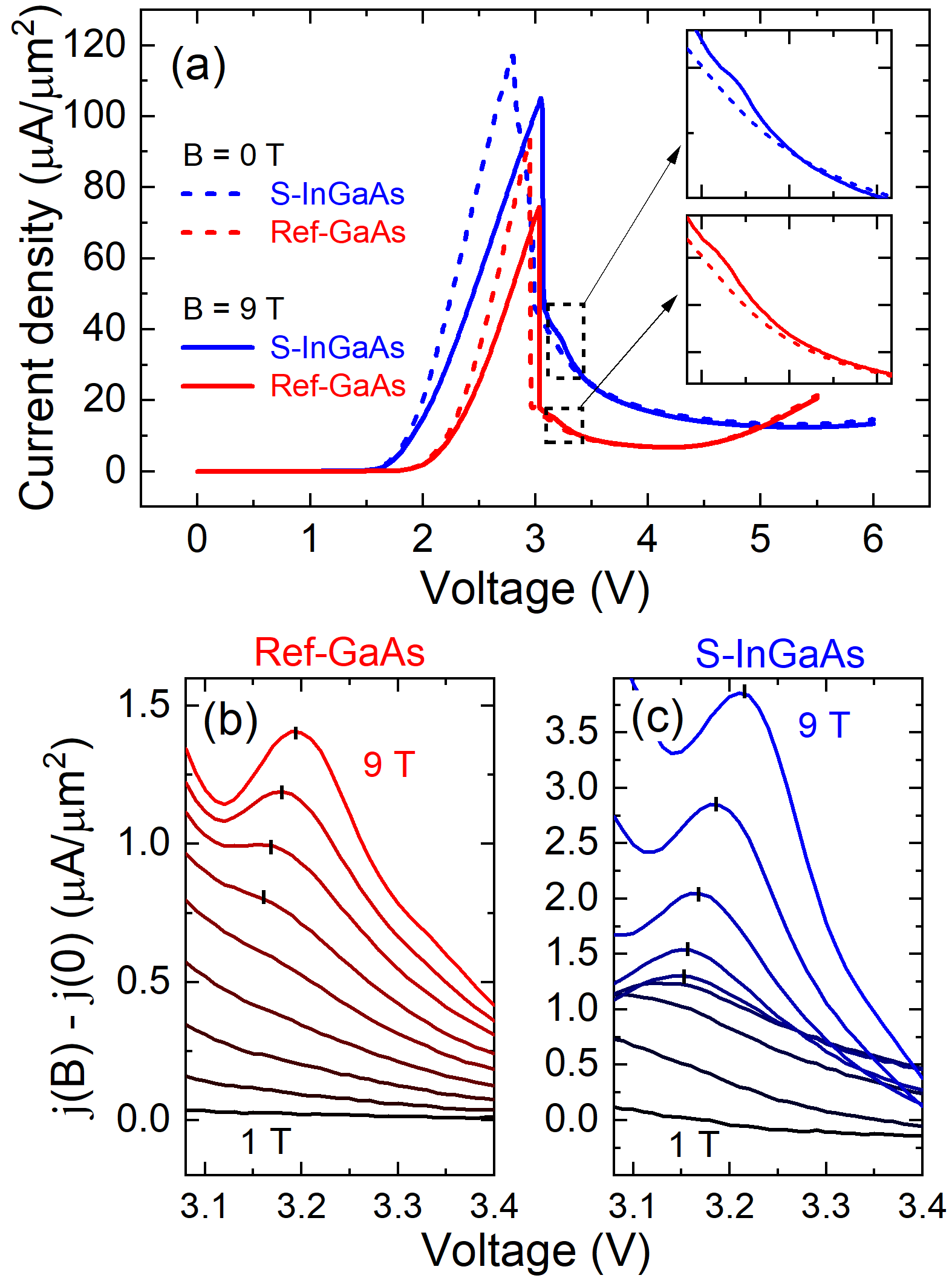}
	\end{center}
	\par
	\vspace{-0.5cm} \caption{(a) Current-voltage characteristics for Ref-GaAs (red) and S-InGaAs (blue) obtained at B = 0 T (dashed lines) and 9 T (solid lines). Insets represent a zoom-in of the voltage range between 3.0 V and 3.8 V. A small shoulder is observed around 3.2 V. Current density difference between curves with applied magnetic field ($j(B)-j(0)$), for (b) Ref-GaAs and (c) S-InGaAs, from 3.08 V to 3.40 V.}
	\label{IV09}
\end{figure}

Figure~\ref{IV09}(a) shows current density characteristics taken at zero magnetic field (dashed lines) and B = 9 T (solid lines) for Ref-GaAs and S-InGaAs colored red and blue, respectively. The S-InGaAs (Ref-GaAs) resonance current density peak decreases from 117 $\mu$A/$\mu$m$^{2}$ (93 $\mu$A/$\mu$m$^{2}$) to 105 $\mu$A/$\mu$m$^{2}$ (74 $\mu$A/$\mu$m$^{2}$) and shifts from 2.80 V (2.95 V) to 3.05 V (3.03 V) by sweeping the magnetic field from 0 to 9 T. PVCR decreases from 8.9 (14.2) to 8.5 (11.2) driven by the reduction in the peak current density. At V = 3.2 V a small shoulder in the current density is observed, as shown in insets of Fig.~\ref{IV09}(a). Figures~\ref{IV09}(b) and ~\ref{IV09}(c) present the current density difference between transport measurements with and without an applied magnetic field (from 1 T to 9 T) in the range between 3.08 and 3.40 V, for Ref-GaAs and S-InGaAs, respectively. Note that a current density peak emerges and changes position with the magnetic field and there is a correlation with the EL intensity increase between 3.1 V and 3.5 V observed in Fig.~\ref{landau}(b). The increased current density with magnetic field leads to a higher hole generation rate and, therefore, the luminescent signal increases. It is important to note that the light emission and current density increase with magnetic field are not proportional. This is because the hole generation rate is enhanced when the system is in the on-resonance regime.~\cite{Cardozo2018} This feature is an evidence of the LL quantization from the DBQW in resonance with the prewell ground state as described in Refs. [\onlinecite{Cylon88,Gobato91,Brown97}].

\begin{figure}[h!]
\par
\begin{center}
\includegraphics[scale=0.35]{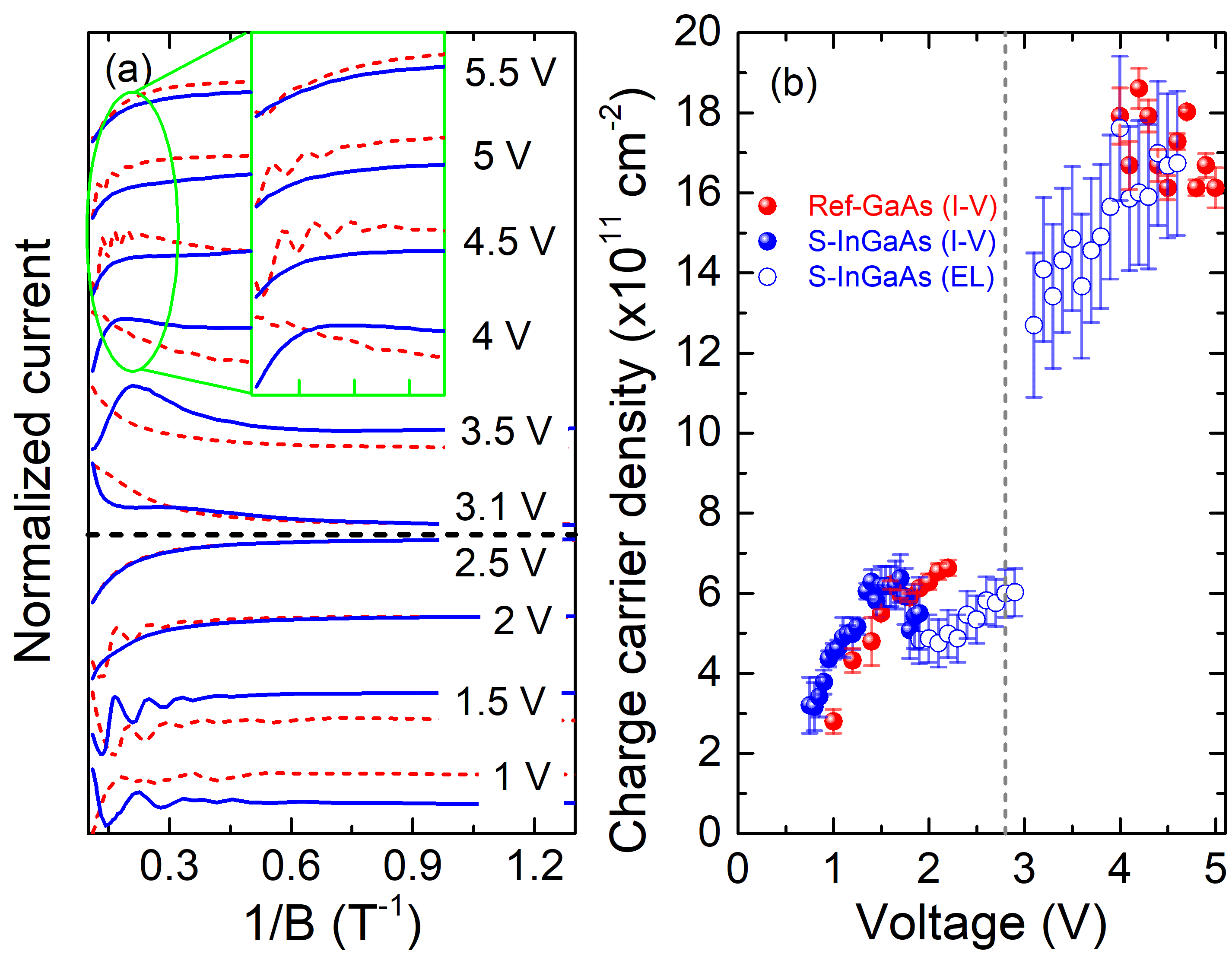}
\end{center}
\par
\vspace{-0.5cm} \caption{(a) Normalized current oscillations as function of magnetic field for several bias voltage values from 1.0 V up to 5.5 V for S-InGaAs (blue lines) and Ref-GaAs (red lines), offseted for clarity. Voltages before (after) the resonance condition are located below (above) the horizontal dashed line. The inset shows a zoom-in for 4.0 V up to 5.5 V. (b) Charge carrier density as function of voltage extracted from the EL emission for S-InGaAs (opened blue circles), and from the current oscillations, for S-InGaAs and Ref-GaAs in blue and red dots, respectively. The vertical dashed line refers to the resonance voltage.}
\label{iv_ib}
\end{figure}

We can extract relevant quantitative information on the charge accumulation by combining both the transport and optical results under magnetic field. First we focus on the current versus magnetic field, from which we can estimate the charge carrier density at the prewell.~\cite{Eaves1989,Goodings94} From the S-InGaAs (Ref-GaAs) I-V characteristics at several magnetic fields, the normalized current oscillations are plotted as function of the inverse field, $1/B$, as blue (red) lines, presented in Fig.~\ref{iv_ib}(a). Once we measured current-voltage characteristics at several magnetic field values, we extracted the current as function of magnetic field by transposing the data and fixing the voltage. Oscillations are visible in both samples, which are the signature of two-dimensional electron gas (2DEG) quantization due to the crossing of the Landau levels and the Fermi energy.~\cite{Goodings94} In the S-InGaAs case, these oscillations are produced by the InGaAs emitter prewell Landau quantization and they can be resolved from 0.75 V to 1.95 V, after which cannot be longer resolved. For the Ref-GaAs sample, an emitter triangular prewell is formed due to the band structure bending with electric field~\cite{Bockenhoff1988}. The oscillations are present from 1 V to 2.2 V, and can be further seen after the resonance, presenting a higher frequency, from 4 V and forward.

Figure~\ref{iv_ib}(b) displays the charge carrier density inside the (triangular) emitter prewell ($\eta_{prewell}$) as function of the voltage for the S-InGaAs (Ref-GaAs) obtained through the current oscillations with period $\Delta(1/B)$~\cite{Eaves1989,Eaves1988,Leadbeater1989} as
\begin{equation}
\eta_{prewell}=\frac{e}{\hbar \pi \Delta(1/B)}.
\end{equation}
Note that, before resonance, the calculated charge carrier densities in both samples are similar, as they were grown with the same donor profile. In the off-resonance case the charge carrier density can be calculated for Ref-GaAs after 4 V, which is observed to be higher than for on-resonance condition due to charge build-up at the emitter barrier.~\cite{Goodings94} It is not possible, however, to calculate the charge carrier density for voltages above the resonance for the S-InGaAs sample using magneto-transport measurements since the oscillations are not observed. The lack of current oscillation after resonance is probably due to the incoherent transport, such as sidewall leakage, thermionic emission, incoherent tunneling, which are not influenced by the magnetic field. For the Ref-GaAs, the oscillations observed from 4 V indicate coherent transport which can be associated to the tunneling through excited DBQW levels. The S-InGaAs coherent transport current oscillations would appear after 6 V, beyond the experimental voltage range. Nevertheless, in this scenario, electroluminescence measurements can be useful for the charge build-up investigation and complete the map of carrier density changes throughout the full operation range of the device. By considering that the charge carriers are thermalized, we can calculate the density of the 2D states at the prewell as follows~\cite{Eaves1989,Eaves1988,Leadbeater1989}
\begin{equation}
\eta_{prewell} = \frac{E_{F}^{prewell}m^{*}}{\hbar^{2}\pi}.
\end{equation}
By analyzing the prewell emission, E$_{1}$, before resonance (Fig.~\ref{landau}(a)), we can define the prewell quasi-Fermi energy, $E_{F}^{prewell}$, as the difference between the peak position and the energy at half maximum, at B = 0 T.~\cite{Libezny94,Lee95} After resonance, as shown in Fig.~\ref{landau}(b), the prewell emission asymmetry interferes with the donor emission line, E$_{2}$, making it difficult to determine the quasi-Fermi energy with the prior procedure. We thus developed an alternative method. By analyzing the EL intensity as function of the magnetic field for fixed energy values, the Landau levels oscillations are clearly resolved. Thus, we determine $E_{F}^{prewell}$ as the difference between E$_{1}$ and the energy position at which oscillations are no longer detected and extrapolate it down to 0 T.

The charge carrier density extracted from the EL measurements using that method is depicted in Fig.~\ref{iv_ib}(b) as open blue circles. The error bars before and after the resonance are estimated considering, respectively, the spectral noise and the donor emission linewidth due to the interference between the LLs and E$_{2}$. As one can see, before resonance these results coincide within the same range of the blue dots, and after resonance, we obtained a charge carrier density similar to the Ref-GaAs sample. These agreements are indications that the optical emission can also be used to estimate these quantitative features.

\section{Conclusions}
In summary, we investigate carrier dynamics of a GaAs/AlGaAs resonant tunneling diode with InGaAs emitter prewell and QW combining transport measurements and electroluminescence. A conventional GaAs/AlGaAs RTD was used as a reference sample, to further study the effects of the S-InGaAs heterostructure. 

The magneto-transport results resemble what has already been reported in the literature. Before the resonant voltage, the charge carrier density obtained for both samples through the current magneto-oscillations are comparable, however the lack of clear oscillations after resonance for the S-InGaAs sample prevents the determination of this parameter. We thus propose an approach to assess the charge accumulation through magneto-electroluminescence where the magneto-transport means are no longer effective. Internal charge carrier transport processes are observed without the need for photoluminescence. This is particularly important as RTDs have been shown to exhibit transport characteristics that can be very sensitive to illumination. Furthermore, the Landau levels quantization in the emitter prewell of a resonant tunneling diode via EL are identified. The spreading of the LLs allowed us to estimate the quasi-Fermi energy at the prewell, and thus, the charge carrier density. The results have shown a good agreement with the transport measurements in the operational range when they are both available. This method of combining complementary transport and optical tools for determining the charge carrier build-up, that controls the operation of photodetectors, might pave the way for developing potentially more efficient devices.

\section{Acknowledgment}

The authors gratefully acknowledge the financial support of the following agencies: Fundação de Amparo à Pesquisa do Estado de São Paulo (FAPESP) (grants $\#$ 2013/18719-1, 2014/19142-1, 2014/02112-3, 2018/01914-0), Conselho Nacional de Desenvolvimento Científico e Tecnológico  (CNPq) (grants $\#$ 163785/2018-0, 471191/2013-2), Coordenação de Aperfeiçoamento de Pessoal de Nível Superior  (CAPES) (grant $\#$88881.133567/2016-01), BAYLAT and the German Ministry of Education, Research (BMBF) within the national project HIRT (FKZ 13XP5003B) and and Photon-N (Grant No. FKZ 13N15125).

\bibliography{bibliography2}

\end{document}